\documentclass{aa}  

\usepackage{graphicx}
\usepackage{txfonts}
\usepackage{color}
\usepackage[hidelinks]{hyperref}
\usepackage{natbib}

\everymath{\displaystyle}

\begin{document}

   \title{A new and simple approach to determine the abundance of hydrogen molecules on interstellar ice mantles}

   \author{U. Hincelin
          \inst{1}
          \and
          Q. Chang\inst{2,3}
          \and
          E. Herbst\inst{1}
          }

   \institute{Department of Chemistry, University of Virginia,
              Charlottesville, VA 22904, USA \\
              \email{ugo.hincelin@virginia.edu}
         \and
             XinJiang Astronomical Observatory,
              Chinese Academy of Sciences, 150 Science 1-Street, Urumqi 830011, PR China
         \and
             Key Laboratory of Radio Astronomy,
               Chinese Academy of Sciences, 2 West Beijing Road, Nanjing 210008, PR China
               }


 
  \abstract
   {Water is usually the main component of ice mantles, which cover the cores of  dust grains in cold portions of dense interstellar clouds.
   When molecular hydrogen is adsorbed onto an icy mantle through physisorption, a common assumption in gas-grain rate-equation models is to use an adsorption energy for molecular hydrogen on a pure water substrate.
   However, at high density and low temperature, when H$_2$ is efficiently adsorbed onto the mantle, its surface abundance can be strongly overestimated if this assumption is still used.
   Unfortunately, the more detailed microscopic Monte Carlo treatment cannot be used to study the abundance of H$_{2}$ in ice mantles if a full gas-grain network is utilized. }
   {We present a numerical method adapted for rate-equation models that takes into account the possibility that an H$_{2}$ molecule can, while diffusing on the surface, find itself bound to another hydrogen molecule, with a far weaker bond than the H$_{2}$-water bond, which can lead to more efficient desorption.  
   We label the ensuing desorption "encounter desorption".}
   {The method is implemented first in a simple system consisting only of hydrogen molecules at steady state between gas and dust using  the rate-equation approach and comparing the results with the results of a microscopic Monte Carlo calculation.
   We then discuss the use of  the rate-equation approach with encounter desorption embedded in a complete gas-grain chemical network.}
   {For the simple system, the rate-equation model with encounter desorption  reproduces the H$_2$ granular coverage computed by the microscopic Monte Carlo model at 10 K for a gas density from $10^4$ to $10^{12}$~cm$^{-3}$, and yields up to a factor 4 difference above $10^{12}$~cm$^{-3}$.
  The H$_2$ granular coverage is also reproduced by a complete gas-grain network.
   We use the rate-equation approach to study the gas-grain chemistry of cold dense regions with and without the encounter desorption mechanism.  
   We find that the grain surface and gas phase species can be sensitive to the H$_2$ coverage, up to several orders of magnitude, depending on the species, the density, and the time considered.}
   {The method is especially useful for dense and cold environments, and for  time-dependent physical conditions, such as occur in  the collapse of dense cores and the formation of protoplanetary disks.
   It is not significantly CPU time consuming, so can be used for example with complex 3D chemical-hydrodynamical simulations.}

   \keywords{astrochemistry --
             molecular processes --
             methods: numerical --
             ISM: molecules --
             ISM: abundances
               }

   \maketitle

\section{Introduction}

Gas-grain chemical models, which are useful tools for studying the chemistry in the interstellar medium, often include the rate-equation approach to calculate the evolution of species in the gas phase and on the grain surface \citep[e.g.,][]{1973ApJ...185..505H,1992ApJS...82..167H}.
Regarding the chemistry on dust grains and their ice mantles, the rate-equation approach can be used in the basic two-phase approach, in which no distinction is made between the surface of the ice and layers underneath it, the three-phase model, in which chemistry only takes place on the surface, or even a multi-layer approach \citep[e.g.,][]{1992ApJS...82..167H,1993MNRAS.263..589H,2012A&A...538A..42T}.
Even the two-phase approach can describe the chemistry reasonably accurately \citep{2008ApJ...682..283G}, and rate-equation methods are efficient for large chemical networks where thousands of reactions are taken into account.
Some limitation exists when the average number of species per dust grain is below unity, and stochastic methods are more accurate \citep{1997OLEB...27...23T,1998ApJ...495..309C,2003Ap&SS.285..725H,2003A&A...400..585L,2004A&A...423..241S}.
Contrary to microscopic kinetic Monte Carlo models, and other detailed stochastic treatments, \citep{2005A&A...434..599C,2005MNRAS.361..565C,2007A&A...469..973C,2009A&A...508..275C,2010A&A...522A..74C,2012ApJ...759..147C,2012ApJ...751...58I,2014ApJ...787..135C}, rate-equation models do not take into account each individual process that occurs on and beneath the ice surface. 
For example, when a molecule is adsorbed on the grain surface, the desorption energy should depend on the substrate and other grain surface parameters, which are functions of the location on the grain surface, itself continuously in evolution as a function of time.
For rate-equation treatments, however, a single binding energy per adsorbate is commonly used depending upon physical conditions and type of source.
Since  water ice is often the main component of dust grain mantles in cold dense interstellar clouds \citep{1982A&A...114..245T,1998ApJ...498L.159W}, the adsorption energies used in models of these sources are usually the ones of a given adsorbate on a water substrate.  A list of desorption energies for a selection of physisorbed adsorbates on water and other substrates can be found in Table 3 of
\cite{2007ApJ...668..294C}.
In this paper we will be concerned with the binding of H$_{2}$ on a water substrate and on itself, and will be using  440~K and  23~K on H$_2$ for the two values, respectively \citep{2007ApJ...668..294C}.

Determination of the amount of H$_{2}$ on a dust grain in a cold cloud is a difficult task for several reasons.  First, it is difficult if not impossible to include the adsorption and desorption of hydrogen molecules in a complete treatment of the surface chemistry via the kinetic Monte Carlo approach given the speed with which these events can happen.  The situation gets worse if complex simulations of star formation, such as three-dimensional hydrodynamical simulations \citep{2012ApJ...758...86F,2013ApJ...775...44H}, or even warm-up models \citep[e.g.][]{2004MNRAS.354.1141V,2006A&A...457..927G}, are undertaken, because the kinetic Monte Carlo models are very time-consuming computationally.  Secondly, the large difference between the binding energy of H$_{2}$ to a water ice substrate and to itself means that a rate-equation model with only one binding energy for H$_{2}$,  the one with water, can lead to
 an overestimate of the H$_2$ granular abundance,  especially at low temperature and high density ($\approx$10~K and >10$^4$~cm$^{-3}$), and on the contrary, considering a single binding energy of  23~K prevents adsorption of H$_2$ onto grain surfaces, which is not real except at high temperatures.  
Thus, a simple and efficient numerical approach to deal with H$_2$ coverage as a function of temperature, density, and time, and applicable to rate-equation chemical models, is desirable.
The goal of this paper is to present one such approach and to use it in treatments of cold and dense regions.  
This new approach differs from earlier approaches of \cite{2010PhRvE..81f1109W}, \cite{2011A&A...529A.151C}, and \cite{2011ApJ...735...15G}.

The remainder of the paper is structured as follows.   
We present our treatment  in terms of a rate equation in Section~\ref{sec:enc_des_mec}.   In Section~\ref{sec:microMCmodel}, we then consider a simple steady-state model in which we only include a fixed amount of H$_{2}$ and calculate the surface abundance of H$_{2}$ as a function of density for a cloud at 10 K.  We compare the results of this simple model with those of a detailed kinetic Monte Carlo approach.  The two approaches lead to very similar results for the H$_{2}$ surface abundance.  In Section~\ref{sec:REmodel},  we introduce a large gas-grain network and code with encounter desorption, based on the Nautilus model \citep{2009A&A...493L..49H}, and use it to obtain the  H$_{2}$ surface abundance as a function of density.  The good agreement with the simple treatments suggests that we can use a large gas-grain rate-equation treatment with encounter desorption to determine the overall chemistry that occurs as a function of H$_2$ surface abundance.  The chemistry is discussed in Section~\ref{sec:discussions}, and a conclusion follows.

\section{The "encounter desorption" mechanism}
\label{sec:enc_des_mec}

Water is the main component of the ice mantle, therefore the desorption energy of a species on a water substrate is usually used.
However, at very high density, H$_2$ can become quite abundant on the grain surface, since it is the most abundant species in the gas phase.  In an extreme case, we would need to use the binding energy of adsorbates to H$_{2}$ and not to water \citep{2013MNRAS.429.3578M}.  
To take this problem into account, \cite{2011ApJ...735...15G} calculated effective binding energies and diffusion barriers according to the fractional coverage of the surface with H$_2$.
This method produces a maximum H$_2$-ice fraction of around $10$~\% under cold molecular cloud conditions.
Unfortunately, desorption energies and diffusion barriers of every species become time dependent using this technique.
Then the differential equation system become stiffer than usual and as a consequence more difficult to solve, which could be a handicap for complex hydrodynamical simulations \citep{2012ApJ...758...86F,2013ApJ...775...44H}, where computing time is a critical limitation.

\begin{figure}
\centering
\includegraphics[width=1.0\linewidth]{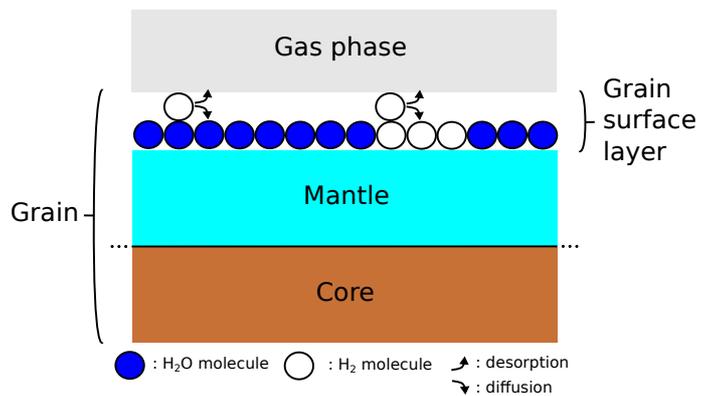}
\caption{An interstellar grain covered by a mantle of ice.  
The surface layer is composed of water and molecular hydrogen, and H$_2$ molecules diffuse on the surface.}
\label{fig:H2_dif_des}
\end{figure}

Our approach, which we label ``encounter'' desorption,  is a different one.   Figure~\ref{fig:H2_dif_des} shows an interstellar grain consisting of a silicate or carbonaceous core, and an ice mantle assumed mainly to be of water ice, with water molecules in the top layer illustrated in dark blue, and hydrogen molecules in white.  Individual hydrogen molecules diffuse for the most part over water molecules until they reach another hydrogen molecule beneath them, at which time the binding energy of the diffusing species is sharply reduced from 440~K to 23~K, raising the likelihood of desorption.
The desorption of an H$_2$  molecule due to the lower desorption energy, when the molecule ends up on an H$_2$ substrate, is modeled by considering the encounter of two H$_2$ molecules on the same surface site.
The grain surface "reaction" $\rm H_2(grain) + H_2(grain) \longrightarrow H_2(grain) + H_2(gas)$ is added to the reaction network with a specific reaction rate $R_{H_2H_2}$ to take into account this process.

The rate is given by
\begin{equation}
\label{eq:enc_des}
R_{H_2H_2}= \frac{1}{2} k_{H_2H_2} n_s(H_2)n_s(H_2)\kappa(H_2),
\end{equation}
in units of cm$^{-3}$~s$^{-1}$, where $k_{H_2H_2}$ is the rate coefficient (cm$^{3}$~s$^{-1}$) at which two hydrogen molecules diffuse into the same lattice site \citep{1992ApJS...82..167H,1998ApJ...495..309C}, the hydrogen concentration on grain surfaces is written as $n_{s}(H_{2})$ (cm$^{-3}$) and $ \kappa(H_{2})$ is the probability of desorption rather than diffusion.
This probability is given by the equation
\begin{equation}
\label{eq:kappa_AdsDes}
\kappa(H_2)=\frac{\displaystyle\sum_{X}k_{Xdes}(H_2)}{R_{diff}(H_2)+\displaystyle\sum_{X}k_{Xdes}(H_2)}
\end{equation}
where the sum over X is a sum over the thermal desorption rate coefficient and assorted non-thermal desorption rate coefficients (s$^{-1}$) due to photons and cosmic rays.
In the general case, we take into account thermal desorption, cosmic ray induced desorption, and photodesorption from direct interstellar UV photons and secondary photons generated by cosmic rays \citep[see][]{1992ApJS...82..167H,1993MNRAS.261...83H,1985A&A...144..147L,2007ApJ...662L..23O,2009A&A...504..891O,2009ApJ...693.1209O,2009A&A...496..281O,2008ApJ...681.1385H,2010A&A...515A..66H}), while $R_{diff}$ is the diffusion rate (s$^{-1}$) of one H$_2$ molecule on an H$_2$ substrate \citep{1992ApJS...82..167H,1998ApJ...495..309C}.   

\section{A comparison between different methods for a simple system}
\label{sec:microMCmodel}
Before we apply our encounter desorption mechanism to a realistic gas-grain simulation, the validity of the mechanism
has to be tested. Since the microscopic Monte Carlo approach is the most rigorous simulation method, ideally
we should compare the results of the rate equation approach including encounter desorption with analogous results with the microscopic Monte Carlo simulation results using a  full reaction network.
However, since the gas phase H$_2$ abundance is too large to be treated by the Monte Carlo method, we 
choose a system that is as simple as possible.   
In this section, we report the results of a comparative study of such a simple system,  in which we consider only H$_{2}$ and a mantle of effectively water ice.  
We then compute the steady-state fraction of molecular hydrogen on the dust grains at 10 K as a function of  total H$_{2}$ density.  
We take a typical grain with radius $0.1 \mu$m and $10^6$ binding sites, and a gas-to-dust number density of 10$^{-12}$. 
H$_2$ from the gas can accrete onto a grain surface and then 
diffuse or desorb from the surface.
Thermal desorption is treated in the standard manner, while the rate coefficient for encounter desorption is treated as in equation (\ref{eq:enc_des}), but without  non-thermal desorption mechanisms. 

The simple rate equation approach is based on setting the time derivative of the concentration of H$_{2}$ on grains to zero:
\begin{equation}
\frac{d~n_{s}(H_2)}{dt} = k_{ads}(H_2)n_g(H_2) - R_{H_2H_2} - k_{\theta des}(H_2)n_s(H_2) = 0 
\end{equation}
and solving for the H$_{2}$ grain concentration.  In the equation,  $n_g(H_{2})$ is the gas-phase H$_{2}$ abundance,  $k_{ads}$ is the adsorption rate coefficient for H$_{2}$, $k_{\theta des}$  is its thermal desorption rate coefficient \citep{1992ApJS...82..167H}, and the rate for encounter desorption is to be found in equation (\ref{eq:enc_des}). 

The microscopic Monte Carlo simulation method has been explained in detail in \cite{2005A&A...434..599C}, so will only be discussed briefly here.
A grain surface with $N$ binding sites is represented as an $L\times L$ square lattice, where $L$ is the number of sites on grain surface in one dimension.
We keep track of the position and movement of H$_{2}$ species on the lattice.
The movements, which include hopping, desorption, and adsorption, are modeled 
as Poisson processes, so that
the time interval between two successive movement operations, $\Delta t$, is given by
\begin{equation}
\Delta t = \frac{\ln(x)}{k},
\end{equation}
where $x$ is a random number uniformly distributed within 0 and 1, and $k$ (in s$^{-1}$) is the hopping rate coefficient $k_{hop}$, the thermal desorption rate coefficient $k_{\theta des}$, or the adsorption rate coefficient $k_{ads}$, depending on the specific movement operation. 
Moreover, hopping will compete with desorption for an H$_{2}$ species that resides in a binding site. 
We combine hopping and desorption as a joint Poisson process and then use a competition mechanism to decide whether the species will hop or desorb \citep{2005A&A...434..599C}.  

\begin{figure}
\centering
\includegraphics[width=1.0\linewidth]{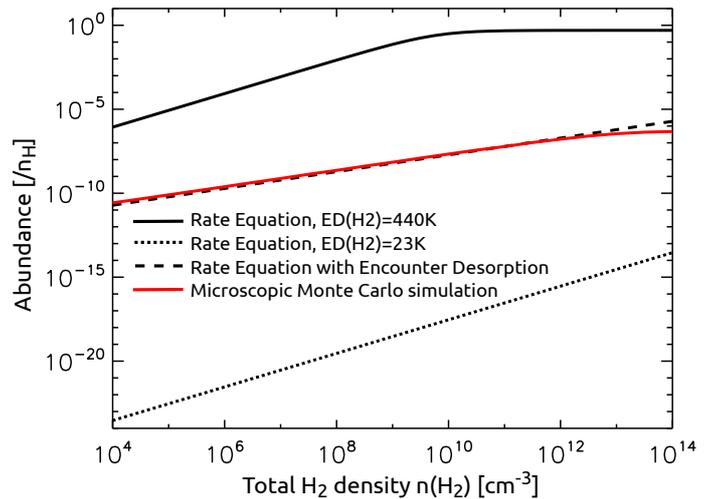}
\caption{H$_2$ fractional abundance on the dust surface with respect to the total proton abundance plotted against the total (gas and grain surface) hydrogen density, as computed by different methods for a simple system at steady state (see text in Section~\ref{sec:microMCmodel}).}
\label{fig:REvsMC}
\end{figure}

Figure~\ref{fig:REvsMC} shows the steady-state molecular hydrogen abundance on a grain surface, calculated as a function of H$_{2}$ total density (gas and grain surface) for three models, two of which contain no encounter desorption using a desorption energy for H$_{2}$ of either 440 K, the H$_{2}$-water value, or 23 K, the H$_{2}$-H$_{2}$ value.
For these models, only the rate-equation result is shown.
The third model contains the encounter desorption rate process as well as thermal desorption using the 440 K desorption energy, which is a very slow process at 10 K\footnote{In this third model, 440~K is used in the rate coefficient $k_{H_2H_2}$ in equation~\ref{eq:enc_des}, while 23~K is used in the different terms of the probability $\kappa(H_2)$ shown in equation~\ref{eq:kappa_AdsDes}.}.
For this case, we also plot the result of the kinetic Monte Carlo approach, which should reproduce the H$_{2}$ granular abundance of the encounter desorption rate-equation model if the latter is accurate.
Note that the kinetic Monte Carlo model assumes a constant gas phase H$_2$ abundance, which is not the case using the rate-equation approach.
The grain surface abundance of H$_2$ is, however, very small compared with the gas phase H$_2$ using the Monte Carlo model, so this assumption does not change our results presented in the figure.

Both models without encounter desorption show a linear dependence of the H$_{2}$ surface abundance on total proton density $n_{\rm H}$ for at least a portion of the H$_{2}$ densities considered.
With the higher desorption energy, H$_{2}$ is slowly desorbed, so that as the H$_{2}$ density approaches 10$^{11}$ cm$^{-3}$, virtually all molecular hydrogen is located on grains, reaching a fractional abundance of 0.5 with respect to the total proton density.  
With the lower desorption energy, the average number of H$_{2}$ molecules per grain is less than unity  even at the highest density utilized (abundance $\approx 10^{-14} - 10^{-13}$).  
With encounter desorption, the results lie in-between, with the H$_{2}$ granular fractional abundance at a standard dense cloud gas density of 10$^{4}$ cm$^{-3}$ approximately 10$^{-9}$, which corresponds to about 40 molecules per grain, and, at the highest density studied, $4\times 10^{6}$ molecules per grain,  which corresponds to $\sim4$ monolayers.  
The surface fractional abundance calculated with the rate equation model  including encounter desorption is slightly larger than the value obtained with the microscopic Monte Carlo model at densities larger than $10^{12}$~cm$^{-3}$, because in the rate-equation model, the H$_2$ grain surface concentration is approximately linearly dependent on the density of the medium, whereas the Monte Carlo model involves one monolayer of H$_2$ as a limit.
However, even at the highest density in our simulation, $10^{14}$~cm$^{-3}$, the encounter desorption model result for the grain H$_{2}$ abundance is only about a factor of 4 larger than the microscopic Monte Carlo model value.

\section{Results with encounter desorption and a large gas-grain reaction network}
\label{sec:REmodel}

Given the degree of agreement between the kinetic Monte Carlo method and the encounter desorption rate-equation method for a simple system, we have chosen to extend the encounter desorption approach to a full gas-grain model, using the Nautilus code \citep{2009A&A...493L..49H}.
The two-phase rate-equation approach is used, in which no distinction is made between the inner and surface layers of the mantle.
Details on the processes included in the code are presented by \cite{2010A&A...522A..42S} and \cite{2012PhDT........49H}.
The potential energy barrier against diffusion, $E_{\rm b}$,  is linked to the desorption energy $E_{\rm D}$ by the equation $E_{\rm b}=\alpha E_{\rm D}$.
We set $\alpha$ equal to 0.5 as in \cite{2006A&A...457..927G}, although other estimates exist, typically ranging from 0.3 to 0.77 \citep{1976RvMP...48..513W,1987ASSL..134..397T,1992ApJS...82..167H,2000MNRAS.319..837R}.
We used the chemical network of \cite{2013ApJ...775...44H}, which includes the latest recommendations from the KIDA experts on gas-phase processes until October 2011.
An electronic version of this network is available at \url{http://kida.obs.u-bordeaux1.fr/models}.
Photodesorption has been included following \cite{2007ApJ...662L..23O,2009A&A...504..891O,2009ApJ...693.1209O,2009A&A...496..281O} and \cite{2008ApJ...681.1385H,2010A&A...515A..66H} and a limiting factor is added to restrict the mechanism to two monolayers.
Two sources of incident UV radiation are considered : direct interstellar UV photons, and secondary photons generated by cosmic rays.
We used the elemental abundances following \cite{2011A&A...530A..61H}\footnote{Values used in this study come from \cite{1982ApJS...48..321G}, \cite{2008ApJ...680..371W}, and \cite{2009ApJ...700.1299J}.} with an oxygen elemental abundance relative to hydrogen of $3.3\times10^{-4}$.
The species are assumed to be initially in an atomic form as in diffuse clouds except for hydrogen, which is initially in H$_2$ form.
Elements with an ionization potential below 13.6~eV -- C, S, Si, Fe, Na, Mg, Cl, and P --  are initially singly ionized.
From the initial state,  the chemistry evolves under cold and dense conditions.
The gas and grain temperature are equal to 10~K, the cosmic-ray ionization rate is $1.3\times10^{-17}$~s$^{-1}$, and the visual extinction is set to 30.
The density is once again varied in the range $\sim 10^{4}$~cm$^{-3}$ to $\sim 10^{14}$~cm$^{-3}$.
We have run three different models, summarized in Table~\ref{tab:models}, which are analogous to those used for the simple models.
In models~440-noED and 23-noED, the desorption energy of H$_2$ is fixed to 440~K and 23~K, respectively, and the encounter desorption mechanism is disabled.
In model~440-ED, the desorption energy of H$_2$ is fixed to 440~K and the encounter desorption mechanism is enabled with a desorption energy equal to 23~K.

\begin{table}[h]
\centering
\caption{Model designations for full gas-grain simulation}
\label{tab:models}
\begin{tabular}{ccc}
Model & H$_2$ Desorption Energy& Encounter Desorption \\
\hline \hline
440-noED & 440~K & disabled \\
23-noED & 23~K & disabled \\
440-ED & 440~K & enabled
\end{tabular}
\end{table}

Figure~\ref{fig:JH2_H2} shows the abundance of H$_2$ at steady state in the gas phase and on a grain surface for all three models, as a function of total gaseous plus surface hydrogen density. 
Steady state for H$_2$ is reached before 10~yr since its high abundance in the gas phase causes a high adsorption rate, and because we start with all hydrogen in its molecular form. The steady-state results in Figure~{\ref{fig:JH2_H2} are quite similar to those in Figure~\ref{fig:REvsMC}.  Thus, the addition of a "complete" gas-grain reaction network does not change significantly the abundance of surface H$_2$ as a function of density.  We note specifically the results for a standard cold dense cloud with the inclusion of encounter desorption (model 440-ED): the H$_{2}$ fractional surface abundance lies between $10^{-11}$ and {\bf $10^{-10}$} for a standard cold dense cloud,  which represent respectively $\sim 10$ and $\sim 100$ molecules on the surface of a dust grain.

The use of encounter desorption, as seen in Figures~\ref{fig:REvsMC}  and \ref{fig:JH2_H2}, clearly reduces the amount of surface H$_2$ at all densities chosen.
These lowered abundances, however, are still significantly higher than what can be obtained if we assume that H$_2$ cannot stick to grains at all, and that all of the molecular hydrogen on grains comes directly from its formation from two H atoms that have accreted onto the surface.
Thus the implementation of encounter desorption does not lead to the same situation as the assumption of no sticking of H$_2$, at least at the densities studied.

The amount of surface hydrogen is likely to affect the chemistry and abundance of other species, both gaseous and solid-state, especially at densities significantly higher than those pertaining to cold dense clouds.  Some of the effect derives from radical-H$_{2}$ surface reactions that can occur even at low temperatures on the surface via tunneling \citep{1993MNRAS.261...83H}.  In the following section,  we discuss the impact of H$_2$ grain coverage on the abundances of other species for sources at 10 K and a range of densities.

\begin{figure}
\centering
\includegraphics[width=1.0\linewidth]{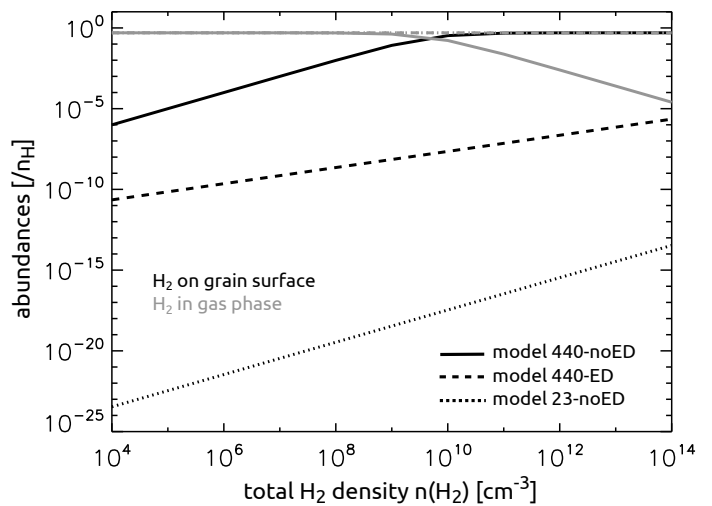}
\caption{H$_2$ fractional abundance in the gas phase (gray) and on the dust grain surface (black) relative to $n_{\rm_H}$ as a function of total H$_{2}$ density for the three models 440-noED (solid line), 23-noED (dotted line), and 440-ED (dashed line).
The dotted and dashed gray lines, which blend into one another, are horizontal and lie atop the figure.}
\label{fig:JH2_H2}
\end{figure}

\section{Discussion}
\label{sec:discussions}

We computed the time-dependent chemical evolution  under the same range of physical conditions as used previously and with the models listed in Table~\ref{tab:models}.  Although the higher density models are not relevant to dense cores, they can be relevant to the dense midplane of protoplanetary disks and to centers of prestellar isothermally collapsing cores.  Moreover, hydrodynamic calculations can lead to temporary high densities and low temperatures.  

We start with a comparison of the  gas-phase abundances measured for the cold cores TMC-1CP and L134N and the gas-phase results of the three models using a comparison parameter $D$ between modeling results and observational constraints, given by the equation
\begin{equation}
D(t)=\frac{1}{N}\sum_j\left|\log\left(X_j^{mod}(t)\right)-\log\left(X_j^{obs}\right)\right|.
\end{equation}
Here, $X_j^{obs}$ is the observed abundance of species $j$, $X_j^{mod}(t)$ is the computed abundance of species $j$ at time $t$, and $N$ is the number of observed species in the cloud.
The smaller the value of $D$, the closer the agreement.
We used the observed abundances listed in \cite{2013ChRv..113.8710A}\footnote{\samepage
Values used in this study come from
\cite{1985ApJ...290..609M}, \cite{1987ApJ...315..646M},
\cite{1989ApJ...345L..63M}, \cite{1991A&A...247..487S},
\cite{1992ApJ...386L..51K}, \cite{1992ApJ...396L..49K},
\cite{1992IAUS..150..171O}, \cite{1993A&A...268..212G},
\cite{1994ApJ...422..621M}, \cite{1994ApJ...427L..51O},
\cite{1997ApJ...486..862P}, \cite{1997ApJ...480L..63L}, \cite{1998A&A...335L...1G},
\cite{1998FaDi..109..205O}, \cite{1998A&A...329.1156T},
\cite{1999ApJ...518..740B}, \cite{2000ApJ...539L.101S},
\cite{2000ApJ...542..870D}, \cite{2000ApJS..126..427T}, \cite{2003A&A...402L..77P},
\cite{2006ApJ...643L..37R}, \cite{2006ApJ...647..412S},
\cite{2007A&A...462..221A}, \cite{2007ApJ...664L..43B},
\cite{2008A&A...478L..19A}, \cite{2009ApJ...690L..27M},
\cite{2009ApJ...691.1494G}, and
\cite{2011A&A...531A.103C}.
}. 
There is little difference in the results for $D(t)$ using the three models at densities of $2 \times 10^{4}$ and $2 \times 10^{5}$~cm$^{-3}$, as shown in Figure~\ref{fig:tmc1_l134n}.

\begin{figure}
\centering
\includegraphics[width=1.0\linewidth]{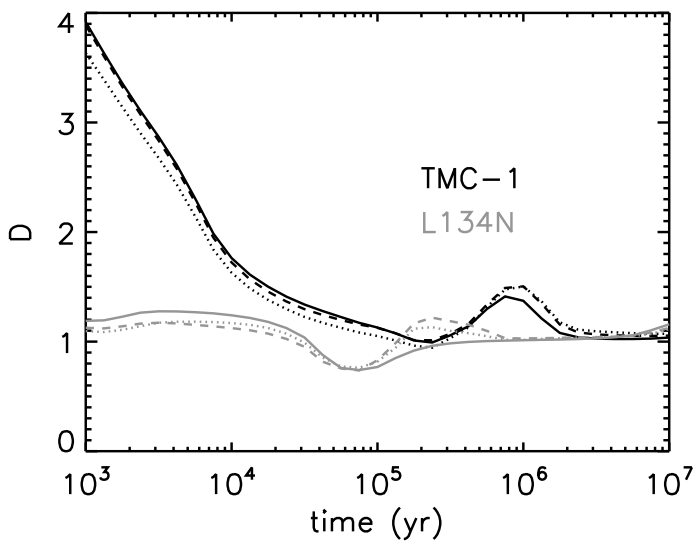}
\caption{Parameter $D$ as a function of time for TMC-1 (black) and L134N (gray), using a total hydrogen density of $2\times10^4$ and $2\times10^5$~cm$^{-3}$ respectively, for the three models: 440-noED (solid line), 23-noED (dotted line), and 440-ED (dashed line).}
\label{fig:tmc1_l134n}
\end{figure}

For individual species, however, the three models can yield different results, even in the gas-phase.
For an example, let us consider the major species water, CO, and methane, and the atomic carbon.
Panels A and B of Figure~\ref{fig:ggH2O_CO_CH4} show the abundances of water, carbon monoxide, and methane both on the grain surface and in the gas phase, as a function of total H$_2$ density, for the three models at $10^6$~yrs, a time relevant for protoplanetary disks and older cold cores.  
Grain surface abundances of these three species are sensitive to the model used, but in the gas phase, only water is strongly affected by the choice of model.  
However, while the grain surface abundances of these species vary by a maximum factor of three, the gas phase water abundance is decreased by three orders of magnitude at $\rm 10^9~cm^{-3}$ going from model 440-ED to 440-noED, which corresponds to an increase in sH$_2$, where the ``s'' stands for ``grain surface''.
The depletion in gaseous H$_{2}$ leads to a depletion of precursors to gaseous water and to an increase of sH$_2$, which consumes sOH so that the production of gas phase water through reactive desorption with sH is lessened.
The abundance of solid atomic carbon, seen in Panel C, also depends strongly on the sH$_{2}$ abundance.
In model 440-noED, where sH$_{2}$ is highest, the abundance of sC is lowest due to its destruction via reaction with sH$_{2}$.

\begin{figure*}
\centering
\includegraphics[width=1.0\linewidth]{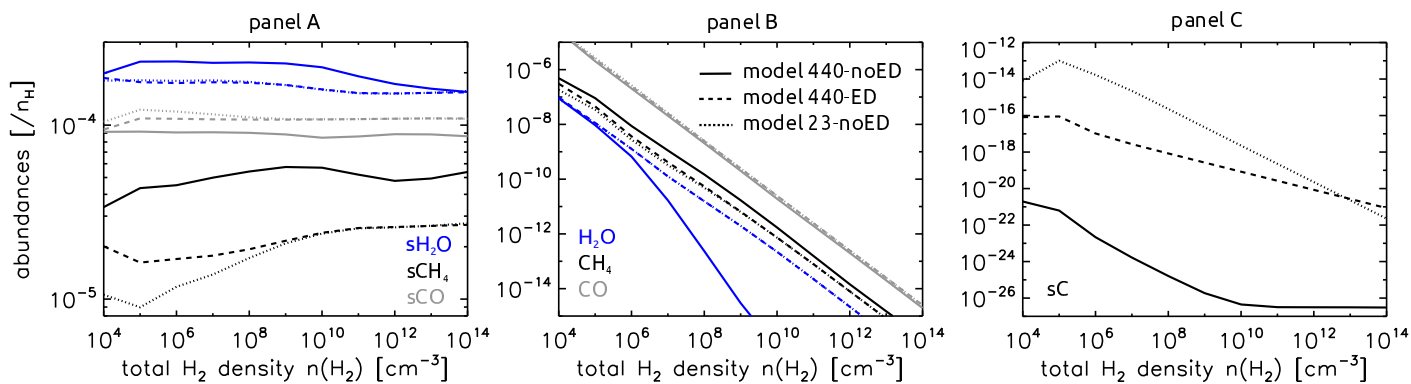}
\caption{H$_2$O (blue), CO (gray), and CH$_4$ (black) abundances at 10$^{6}$ yr on a grain surface (panel A) and in the gas phase (panel B)  plotted against total H$_{2}$ density for the three models 440-noED (solid line), 23-noED (dotted line), and 440-ED (dashed line).
Panel C contains the atomic carbon abundance on a grain surface at the same time and the same models.}
\label{fig:ggH2O_CO_CH4}
\end{figure*}

Depending on the density and the time considered,  a general behavior can be seen for the majority of grain surface species, based upon the surface H$_{2}$ abundance (see Figure~\ref{fig:JH2_H2}).
At the lowest densities, model 440-noED and 440-ED present the same results, while the results of model 23-noED are different.  
At the highest densities, model 440-noED shows different results from the two others, which are quite similar.
Thus the encounter desorption model starts out similarly to the model with a desorption energy for H$_2$ of 440~K and ends up, with increasing density, similar to the model with a lower desorption energy of 23~K.
This relation can be understood by the following argument.
At the lowest densities, H$_2$ does not adsorb very much, so encounter desorption is not very efficient since sH$_2$ lies mainly on top of the water substrate.   
At the highest densities, H$_2$ builds hundreds or even thousands of monolayers if we consider a fixed desorption energy of H$_2$ of 440 K, while encounter desorption becomes quite efficient in these conditions.
We can then discriminate among three different regimes.
The first one occurs when almost no H$_2$ at all is present on the grain surface (model 23-noED at the lowest densities), the second one when some H$_2$ is present at an "intermediate level" (models 440-ED and 440-noED  at the lowest densities, and models 440-ED and 23-noED at the highest densities), and a third one when H$_2$ is very abundant on the grain surface and depletion of H$_2$ from the gas phase is large (model 440-noED at the highest densities).
 
Figure~\ref{fig:JCO2} shows these different regimes and the transition from one regime to another one for sCO$_2$.
At $2\times 10^5$~yr and $10^4$~cm$^{-3}$, models 440-noED and 440-ED give close results while model 23-noED gives $\sim 4$ times more sCO$_2$.
Between $10^5$ and $10^{11}$~cm$^{-3}$, each model gives different results, during this transitional range.
At the highest densities, models 440-ED and 23-noED give same result and the third model gives slightly different results.
However, these transitions are not only density dependent, but also somewhat time dependent.
At $10^7$~yrs and $10^4$~cm$^{-3}$, all models give similar results, while outside this density range, models 440-ED and 23-noED give similar results and model 440-noED gives different ones.

\begin{figure}
\centering
\includegraphics[width=1.0\linewidth]{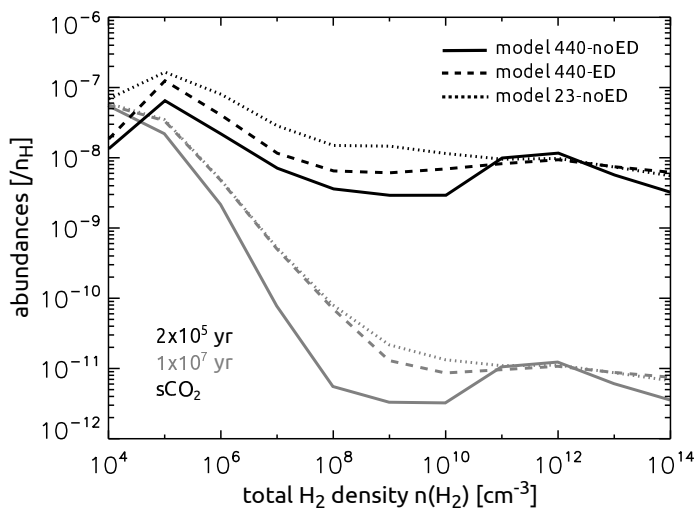}
\caption{CO$_2$ abundance on the grain surface at $2\times 10^5$~yrs (black) and $10^7$~yrs (gray) for the three models 440-noED (solid line), 23-noED (dotted line), and 440-ED (dashed line).}
\label{fig:JCO2}
\end{figure}

\subsection{Sticking probability}

We assume a sticking probability $S$ equal to unity, which means that each collision between a gas phase species and a grain results in an adsorption.
In reality, this probability depends on numerous parameters such as the gas and grain temperature \citep[e.g.][]{2011EPJWC..1803002F} as well as the composition and the structure of the grain surface \citep[e.g.][]{1970JChPh..53...79H,1991ApJ...379..647B,1998A&A...330..773M}.
Recently, \cite{2014MNRAS.443.1301A} has experimentally studied the sticking coefficient of H$_2$ on an olivine substrate, and estimated a lower limit from 0.25 to 0.82 for temperatures between 7~K and 14~K.
Our initial assumption may have some impact on our results, so we tested the sensitivity of the rate equation model to this coefficient using a value equal to 0.5. 

With $S=0.5$,  the abundance of sH$_2$ is decreased while the abundance of gas phase H$_2$ is increased, both by a factor $\sim 10$ maximum.
In the 440-noED model, all hydrogen is located on the grain surface as the densities approaches $10^{12}$~cm$^{-3}$, a roughly ten times higher value than with $S=1$.
This result comes obviously from the lower adsorption rate of gas phase H$_2$ to the grain surface.
The sensitivity of other molecules to the value of $S$ also comes from whether they are formed in the gas or on grains.
Molecules such sH$_2$O and sCH$_4$ are efficiently produced on the grain surface, so a decrease in the adsorption rate normally implies a decrease of the abundance of the reactants that will produce these molecules.
However, water is also formed in the gas phase , so its sensitivity to $S$ is lower than for methane which is essentially formed on the grain surface; the abundance of water is decreased by a factor of two at most while the factor can be as high as ten for methane.
The abundance of sCO is  modified by less than a factor two. 
This small effect stems from two opposing processes: the adsorption rate of CO is lower when $S$ is reduced to 0.5, but the surface reaction rates involving sCO are also lower. 

While the abundance for a given molecule can be different when the sticking probability is reduced from 1.0 to 0.5,  the relative results of the three models 440-ED, 440-noED, and 23-noED exhibit the same pattern whatever the value of the coefficient is.
For example, the abundance of sC is still much lower in the case of the 440-noED model than the other two models, while the gas phase abundances of CO, CH$_4$, and H$_2$O are still not dependent on the model, except for water at densities higher than $10^7$~cm$^{-3}$ for the 440-noED model as shown in panel B of figure~\ref{fig:ggH2O_CO_CH4}.
As a consequence, we conclude that the value of the sticking coefficient does not impact the relative efficiency of encounter desorption significantly.

\subsection{Initial abundances}

We typically start with all hydrogen in its molecular form.
To test the sensitivity of encounter desorption to this assumption, we also performed some simulations with all hydrogen initially in its pure atomic form.
The results for the three models 440-ED, 440-noED, and 23-noED are presented in Figure~\ref{fig:ggH2O_CO_CH4_HinH}, at the same time (10$^{6}$ yr) and for the same species as in Figure~\ref{fig:ggH2O_CO_CH4} to allow for an easy comparison.
The abundance profiles of surface atomic carbon and gas phase water, carbon monoxide, and methane are similar to our previous simulation.
Due to the high density, adsorption of atomic hydrogen is efficient, and H$_2$ is formed quickly on the grain surface.
Grain surface H$_2$ needs about 1~yr or less depending on the density to reach steady state, and gas phase H$_2$ needs about $10^6$ and $10^1$~yrs to reach steady state at a total proton density of $2\times 10^4$ and $2\times 10^8$~cm$^{-3}$ respectively.
The case of the main ices is however slightly different.
They are formed faster, since hydrogenation by s-H is more efficient.
Besides, surface atomic carbon is primarily converted into methane rather than carbon monoxide.
For these ices, the abundance of sH$_2$ seems less critical than for our previous simulations and as a consequence differences between the results of models 440-ED, 440-noED, and 23-noED are reduced.
In conclusion, our results are still sensitive to encounter desorption at 10$^{6}$ yr using atomic hydrogen as an initial condition, depending on the considered species.

\begin{figure*}
\centering
\includegraphics[width=1.0\linewidth]{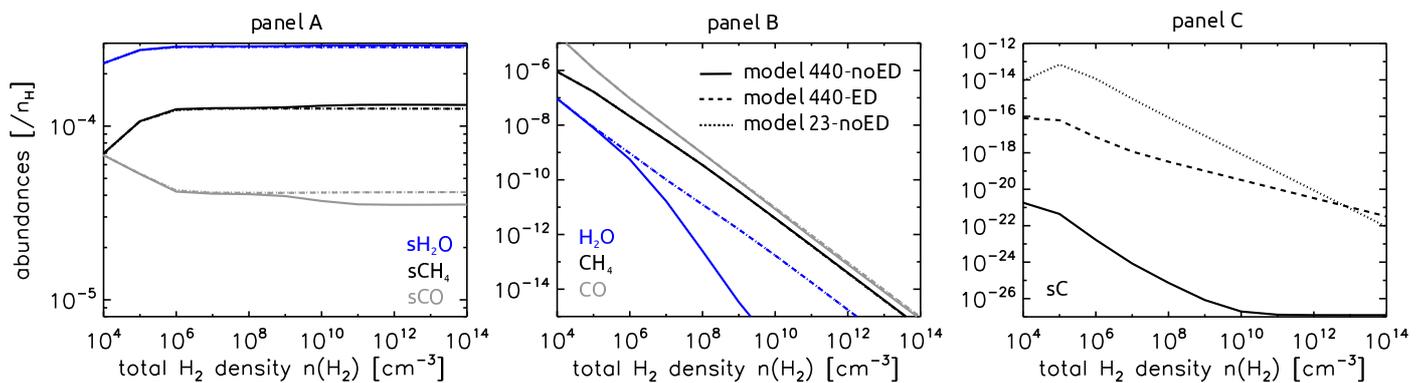}
\caption{Same as Figure~\ref{fig:ggH2O_CO_CH4}. Hydrogen is initially in atomic rather than molecular form.}
\label{fig:ggH2O_CO_CH4_HinH}
\end{figure*}

\subsection{Motion through quantum tunneling}

We typically assume thermal diffusion.
We also studied the sensitivity of our three models 440-ED, 440-noED, and 23-noED to the motion of H$_2$ through quantum tunneling.
The abundances of H$_2$ in the gas phase and on the grain surface are not changed for the two models 440-noED and 23-noED using this new assumption, at all times and densities.
For the third model, abundance of s-H$_2$ is however reduced by about three orders of magnitude compared to the same model without tunneling, at all times and densities.
Since motion through quantum tunneling is faster than thermal diffusion at 10~K, encounter desorption happens more frequently and reduces the surface abundance of H$_2$.
As a consequence for the molecules studied in this paper, the results of 440-ED model are closer to those of the 23-noED model.
Depending on the density and the time however, results of these two models can still be quite different.

\section{Conclusion}
\label{sec:conclu}

We have developed a new approach to prevent a huge accumulation of H$_2$ on interstellar grain surfaces at low temperatures and high densities, which should not occur because the desorption energy of H$_2$ on an H$_2$ substrate is much lower than on a water substrate.
This method, which is to be used in gas-grain rate-equation simulations,  is based on the facile desorption of molecular hydrogen when it encounters a molecule of an H$_2$ substrate.
We have named this process ``encounter desorption''.

In order to test our approach, we first used a very simple system including the encounter desorption process to calculate the surface abundance of H$_{2}$ molecules as a function of density at a temperature of 10 K.  We then compared our result with an analogous but more exact result obtained using a microscopic Monte Carlo stochastic method.  A comparison between the steady-state results of the rate-equation model with encounter desorption and the  Monte-Carlo approach gives very good agreement for gas densities from $10^4$ to $10^{12}$~cm$^{-3}$.
Above $10^{12}$~cm$^{-3}$, the rate-equation model slightly overestimates the H$_2$ grain surface abundance, by a factor of up to 4.
We then used a complete gas-grain network with the Nautilus model to which encounter desorption was added and repeated the steady-state rate-equation calculation of the surface abundance of H$_{2}$ vs density, obtaining very similar results to both the simple rate-equation and Monte Carlo treatments.  We thus conclude that the approach is a reasonable one, although the mathematics do not distinguish between the surface layer of an ice mantle and the inner layers.  

We also studied the impact of different H$_2$ surface abundances on the abundance of other species, using two models with a fixed desorption energy of H$_2$ (23~K and 440~K) and one model with the encounter desorption mechanism. The sensitivity of the results of these models is relatively complex, and depends on the considered chemical species, the density of the medium, and the time.  Nevertheless, the calculated abundances can often be divided into three regions depending upon whether the H$_{2}$ surface coverage is low, intermediate, or high.  
Reducing the sticking coefficient for adsorption from unity to 0.5 does not change these results.

We tested the sensitivity of our models to a different initial condition -- hydrogen initially in atomic form instead of molecular -- and to the motion of H$_2$ via quantum tunneling instead of only thermal diffusion.
These assumptions may change the abundances of some species depending on the density and the time, but our simulations are generally still sensitive to the encounter desorption mechanism, and  the results mentioned in the previous paragraph often still hold.

Our results highlight the need to incorporate in rate-equation models a way to more correctly model H$_2$ coverage at low temperature and high density.
This need  becomes even more pressing when physical conditions present huge variations, such as during the collapse of a prestellar core to form a protostar surrounded by a disk.
In this scenario, matter transits through variations of both density and temperature, which lead to potentially large variations of H$_2$ coverage, some quite unphysical such as the limit in which all H$_{2}$ lies on interstellar grain mantles.
The encounter desorption method keeps the amount of H$_{2}$ at physically reasonable values by using both the desorption energy of H$_{2}$ on a water ice mantle and on a mantle with some H$_{2}$ on its surface.  The approach computes the rate of encounter desorption ``on the fly'', i.e.,  as a function of time and H$_2$ coverage, and therefore is well designed for a scenario in which the physical conditions are rapidly changing.  
In addition, it is relatively easy to implement, and is not significantly CPU time consuming.
Although H$_2$ is the most abundant molecule in the interstellar medium, a possible extension to this work would be to consider encounter desorption for other weakly bound atoms and molecules.

\begin{acknowledgements}
U. Hincelin thanks G. Hassel for help in adding photodesorption processes to the chemical code, and K. Acharyya and K. Furuya for useful discussions.
We thank the anonymous referee for suggestions that helped us to improve our paper.
E. Herbst acknowledges the support of the National Science Foundation (US) for his astrochemistry program, and  support from the NASA Exobiology and Evolutionary Biology program through a subcontract from Rensselaer Polytechnic Institute.
Some kinetic data we used have been downloaded from the online database KIDA (KInetic Database for Astrochemistry, \url{http://kida.obs.u-bordeaux1.fr}, \cite{2012ApJS..199...21W}).
\end{acknowledgements}


\begin{thebibliography}{}
\bibitem[Acharyya(2014)]{2014MNRAS.443.1301A} Acharyya, K.\ 2014, \mnras, 443, 1301
\bibitem[Ag{\'u}ndez \& Wakelam(2013)]{2013ChRv..113.8710A} Ag{\'u}ndez, M., \& Wakelam, V.\ 2013, Chemical Reviews, 113, 8710
\bibitem[Ag{\'u}ndez et al.(2008)]{2008A&A...478L..19A} Ag{\'u}ndez, M., Cernicharo, J., Gu{\'e}lin, M., et al.\ 2008, \aap, 478, L19
\bibitem[Akyilmaz et al.(2007)]{2007A&A...462..221A} Akyilmaz, M., Flower, D.~R., Hily-Blant, P., Pineau Des For{\^e}ts, G., \& Walmsley, C.~M.\ 2007, \aap, 462, 221 
\bibitem[Bell et al.(1999)]{1999ApJ...518..740B} Bell, M.~B., Feldman, P.~A., Watson, J.~K.~G., et al.\ 1999, \apj, 518, 740
\bibitem[Br{\"u}nken et al.(2007)]{2007ApJ...664L..43B} Br{\"u}nken, S., Gupta, H., Gottlieb, C.~A., McCarthy, M.~C., \& Thaddeus, P.\ 2007, \apjl, 664, L43 
\bibitem[Buch \& Zhang(1991)]{1991ApJ...379..647B} Buch, V., \& Zhang, Q.\ 1991, \apj, 379, 647
\bibitem[Caselli et al.(1998)]{1998ApJ...495..309C} Caselli, P., Hasegawa, T.~I., \& Herbst, E.\ 1998, \apj, 495, 309
\bibitem[Cazaux et al.(2010)]{2010A&A...522A..74C} Cazaux, S., Cobut, V., Marseille, M., Spaans, M., \& Caselli, P.\ 2010, \aap, 522, A74
\bibitem[Cernicharo et al.(2011)]{2011A&A...531A.103C} Cernicharo, J., Spielfiedel, A., Balan{\c c}a, C., et al.\ 2011, \aap, 531, A103
\bibitem[Chang \& Herbst(2012)]{2012ApJ...759..147C} Chang, Q., \& Herbst, E.\ 2012, \apj, 759, 147
\bibitem[Chang \& Herbst(2014)]{2014ApJ...787..135C} Chang, Q., \& Herbst, E.\ 2014, \apj, 787, 135
\bibitem[Chang et al.(2005)]{2005A&A...434..599C} Chang, Q., Cuppen, H.~M., \& Herbst, E.\ 2005, \aap, 434, 599
\bibitem[Chang et al.(2007)]{2007A&A...469..973C} Chang, Q., Cuppen, H.~M., \& Herbst, E.\ 2007, \aap, 469, 973
\bibitem[Cuppen \& Garrod(2011)]{2011A&A...529A.151C} Cuppen, H.~M., \& Garrod, R.~T.\ 2011, \aap, 529, A151
\bibitem[Cuppen \& Herbst(2005)]{2005MNRAS.361..565C} Cuppen, H.~M., \& Herbst, E.\ 2005, \mnras, 361, 565
\bibitem[Cuppen \& Herbst(2007)]{2007ApJ...668..294C} Cuppen, H.~M., \& Herbst, E.\ 2007, \apj, 668, 294
\bibitem[Cuppen et al.(2009)]{2009A&A...508..275C} Cuppen, H.~M., van Dishoeck, E.~F., Herbst, E., \& Tielens, A.~G.~G.~M.\ 2009, \aap, 508, 275
\bibitem[Dickens et al.(2000)]{2000ApJ...542..870D} Dickens, J.~E., Irvine, W.~M., Snell, R.~L., et al.\ 2000, \apj, 542, 870
\bibitem[Fillion et al.(2011)]{2011EPJWC..1803002F} Fillion, J.~H., Dulieu, F., Romanzin, C., \& Cazaux, S.\ 2011, European Physical Journal Web of Conferences, 18, 3002
\bibitem[Furuya et al.(2012)]{2012ApJ...758...86F} Furuya, K., Aikawa, Y., Tomida, K., et al.\ 2012, \apj, 758, 86
\bibitem[Garrod \& Herbst(2006)]{2006A&A...457..927G} Garrod, R.~T., \& Herbst, E.\ 2006, \aap, 457, 927
\bibitem[Garrod \& Pauly(2011)]{2011ApJ...735...15G} Garrod, R.~T., \& Pauly, T.\ 2011, \apj, 735, 15
\bibitem[Garrod et al.(2008)]{2008ApJ...682..283G} Garrod, R.~T., Weaver, S.~L.~W., \& Herbst, E.\ 2008, \apj, 682, 283
\bibitem[Gerin et al.(1993)]{1993A&A...268..212G} Gerin, M., Viala, Y., \& Casoli, F.\ 1993, \aap, 268, 212
\bibitem[Graedel et al.(1982)]{1982ApJS...48..321G} Graedel, T.~E., Langer, W.~D., \& Frerking, M.~A.\ 1982, \apjs, 48, 321
\bibitem[Guelin et al.(1998)]{1998A&A...335L...1G} Guelin, M., Neininger, N., \& Cernicharo, J.\ 1998, \aap, 335, L1
\bibitem[Gupta et al.(2009)]{2009ApJ...691.1494G} Gupta, H., Gottlieb, C.~A., McCarthy, M.~C., \& Thaddeus, P.\ 2009, \apj, 691, 1494
\bibitem[Hasegawa \& Herbst(1993a)]{1993MNRAS.261...83H} Hasegawa, T.~I., \& Herbst, E.\ 1993a, \mnras, 261, 83
\bibitem[Hasegawa \& Herbst(1993b)]{1993MNRAS.263..589H} Hasegawa, T.~I., \& Herbst, E.\ 1993b, \mnras, 263, 589 
\bibitem[Hasegawa et al.(1992)]{1992ApJS...82..167H} Hasegawa, T.~I., Herbst, E., \& Leung, C.~M.\ 1992, \apjs, 82, 167
\bibitem[Hassel et al.(2008)]{2008ApJ...681.1385H} Hassel, G.~E., Herbst, E., \& Garrod, R.~T.\ 2008, \apj, 681, 1385
\bibitem[Hassel et al.(2010)]{2010A&A...515A..66H} Hassel, G.~E., Herbst, E., \& Bergin, E.~A.\ 2010, \aap, 515, A66 
\bibitem[Herbst \& Klemperer(1973)]{1973ApJ...185..505H} Herbst, E., \& Klemperer, W.\ 1973, \apj, 185, 505
\bibitem[Herbst \& Shematovich(2003)]{2003Ap&SS.285..725H} Herbst, E., \& Shematovich, V.~I.\ 2003, \apss, 285, 725
\bibitem[Hersant et al.(2009)]{2009A&A...493L..49H} Hersant, F., Wakelam, V., Dutrey, A., Guilloteau, S., \& Herbst, E.\ 2009, \aap, 493, L49
\bibitem[Hincelin(2012)]{2012PhDT........49H} Hincelin, U.\ 2012, Ph.D.~Thesis
\bibitem[Hincelin et al.(2011)]{2011A&A...530A..61H} Hincelin, U., Wakelam, V., Hersant, F., et al.\ 2011, \aap, 530, A61
\bibitem[Hincelin et al.(2013)]{2013ApJ...775...44H} Hincelin, U., Wakelam, V., Commer{\c c}on, B., Hersant, F., \& Guilloteau, S.\ 2013, \apj, 775, 44
\bibitem[Hollenbach \& Salpeter(1970)]{1970JChPh..53...79H} Hollenbach, D., \& Salpeter, E.~E.\ 1970, \jcp, 53, 79
\bibitem[Iqbal et al.(2012)]{2012ApJ...751...58I} Iqbal, W., Acharyya, K., \& Herbst, E.\ 2012, \apj, 751, 58
\bibitem[Jenkins(2009)]{2009ApJ...700.1299J} Jenkins, E.~B.\ 2009, \apj, 700, 1299
\bibitem[Kawaguchi et al.(1992a)]{1992ApJ...386L..51K} Kawaguchi, K., Ohishi, M., Ishikawa, S.-I., \& Kaifu, N.\ 1992a, \apjl, 386, L51
\bibitem[Kawaguchi et al.(1992b)]{1992ApJ...396L..49K} Kawaguchi, K., Takano, S., Ohishi, M., et al.\ 1992b, \apjl, 396, L49
\bibitem[Langer et al.(1997)]{1997ApJ...480L..63L} Langer, W.~D., Velusamy, T., Kuiper, T.~B.~H., et al.\ 1997, \apjl, 480, L63
\bibitem[Leger et al.(1985)]{1985A&A...144..147L} Leger, A., Jura, M., \& Omont, A.\ 1985, \aap, 144, 147
\bibitem[Lipshtat \& Biham(2003)]{2003A&A...400..585L} Lipshtat, A., \& Biham, O.\ 2003, \aap, 400, 585
\bibitem[Marcelino et al.(2009)]{2009ApJ...690L..27M} Marcelino, N., Cernicharo, J., Tercero, B., \& Roueff, E.\ 2009, \apjl, 690, L27
\bibitem[Masuda et al.(1998)]{1998A&A...330..773M} Masuda, K., Takahashi, J., \& Mukai, T.\ 1998, \aap, 330, 773
\bibitem[Matthews et al.(1985)]{1985ApJ...290..609M} Matthews, H.~E., Friberg, P., \& Irvine, W.~M.\ 1985, \apj, 290, 609
\bibitem[Matthews et al.(1987)]{1987ApJ...315..646M} Matthews, H.~E., MacLeod, J.~M., Broten, N.~W., Madden, S.~C., \& Friberg, P.\ 1987, \apj, 315, 646
\bibitem[McGonagle et al.(1994)]{1994ApJ...422..621M} McGonagle, D., Irvine, W.~M., \& Ohishi, M.\ 1994, \apj, 422, 621
\bibitem[Minh et al.(1989)]{1989ApJ...345L..63M} Minh, Y.~C., Irvine, W.~M., \& Ziurys, L.~M.\ 1989, \apjl, 345, L63
\bibitem[Morata \& Hasegawa(2013)]{2013MNRAS.429.3578M} Morata, O., \& Hasegawa, T.~I.\ 2013, \mnras, 429, 3578
\bibitem[{\"O}berg et al.(2007)]{2007ApJ...662L..23O} {\"O}berg, K.~I., Fuchs, G.~W., Awad, Z., et al.\ 2007, \apjl, 662, L23
\bibitem[{\"O}berg et al.(2009a)]{2009A&A...504..891O} {\"O}berg, K.~I., Garrod, R.~T., van Dishoeck, E.~F., \& Linnartz, H.\ 2009a, \aap, 504, 891 
\bibitem[{\"O}berg et al.(2009b)]{2009ApJ...693.1209O} {\"O}berg, K.~I., Linnartz, H., Visser, R., \& van Dishoeck, E.~F.\ 2009b, \apj, 693, 1209 
\bibitem[{\"O}berg et al.(2009c)]{2009A&A...496..281O} {\"O}berg, K.~I., van Dishoeck, E.~F., \& Linnartz, H.\ 2009c, \aap, 496, 281 
\bibitem[Ohishi \& Kaifu(1998)]{1998FaDi..109..205O} Ohishi, M., \& Kaifu, N.\ 1998, Faraday Discussions, 109, 205
\bibitem[Ohishi et al.(1992)]{1992IAUS..150..171O} Ohishi, M., Irvine, W.~M., \& Kaifu, N.\ 1992, Astrochemistry of Cosmic Phenomena, 150, 171
\bibitem[Ohishi et al.(1994)]{1994ApJ...427L..51O} Ohishi, M., McGonagle, D., Irvine, W.~M., Yamamoto, S., \& Saito, S.\ 1994, \apjl, 427, L51
\bibitem[Pagani et al.(2003)]{2003A&A...402L..77P} Pagani, L., Olofsson, A.~O.~H., Bergman, P., et al.\ 2003, \aap, 402, L77
\bibitem[Pratap et al.(1997)]{1997ApJ...486..862P} Pratap, P., Dickens, J.~E., Snell, R.~L., et al.\ 1997, \apj, 486, 862
\bibitem[Remijan et al.(2006)]{2006ApJ...643L..37R} Remijan, A.~J., Hollis, J.~M., Snyder, L.~E., Jewell, P.~R., \& Lovas, F.~J.\ 2006, \apjl, 643, L37
\bibitem[Ruffle \& Herbst(2000)]{2000MNRAS.319..837R} Ruffle, D.~P., \& Herbst, E.\ 2000, \mnras, 319, 837
\bibitem[Semenov et al.(2010)]{2010A&A...522A..42S} Semenov, D., Hersant, F., Wakelam, V., et al.\ 2010, \aap, 522, A42
\bibitem[Schilke et al.(1991)]{1991A&A...247..487S} Schilke, P., Walmsley, C.~M., Henkel, C., \& Millar, T.~J.\ 1991, \aap, 247, 487
\bibitem[Snell et al.(2000)]{2000ApJ...539L.101S} Snell, R.~L., Howe, J.~E., Ashby, M.~L.~N., et al.\ 2000, \apjl, 539, L101
\bibitem[Snyder et al.(2006)]{2006ApJ...647..412S} Snyder, L.~E., Hollis, J.~M., Jewell, P.~R., Lovas, F.~J., \& Remijan, A.\ 2006, \apj, 647, 412
\bibitem[Stantcheva \& Herbst(2004)]{2004A&A...423..241S} Stantcheva, T., \& Herbst, E.\ 2004, \aap, 423, 241
\bibitem[Takano et al.(1998)]{1998A&A...329.1156T} Takano, S., Masuda, A., Hirahara, Y., et al.\ 1998, \aap, 329, 1156
\bibitem[Taquet et al.(2012)]{2012A&A...538A..42T} Taquet, V., Ceccarelli, C., \& Kahane, C.\ 2012, \aap, 538, A42 
\bibitem[Tielens \& Allamandola(1987)]{1987ASSL..134..397T} Tielens, A.~G.~G.~M., \& Allamandola, L.~J.\ 1987, Interstellar Processes, 134, 397 
\bibitem[Tielens \& Charnley(1997)]{1997OLEB...27...23T} Tielens, A.~G.~G.~M., \& Charnley, S.~B.\ 1997, Origins of Life and Evolution of the Biosphere, 27, 23
\bibitem[Tielens \& Hagen(1982)]{1982A&A...114..245T} Tielens, A.~G.~G.~M., \& Hagen, W.\ 1982, \aap, 114, 245
\bibitem[Turner et al.(2000)]{2000ApJS..126..427T} Turner, B.~E., Herbst, E., \& Terzieva, R.\ 2000, \apjs, 126, 427 
\bibitem[Viti et al.(2004)]{2004MNRAS.354.1141V} Viti, S., Collings, M.~P., Dever, J.~W., McCoustra, M.~R.~S., \& Williams, D.~A.\ 2004, \mnras, 354, 1141
\bibitem[Wakelam \& Herbst(2008)]{2008ApJ...680..371W} Wakelam, V., \& Herbst, E.\ 2008, \apj, 680, 371
\bibitem[Wakelam et al.(2012)]{2012ApJS..199...21W} Wakelam, V., Herbst, E., Loison, J.-C., et al.\ 2012, \apjs, 199, 21
\bibitem[Watson(1976)]{1976RvMP...48..513W} Watson, W.~D.\ 1976, Reviews of Modern Physics, 48, 513
\bibitem[Wolff et al.(2010)]{2010PhRvE..81f1109W} Wolff, A., Lohmar, I., Krug, J., Frank, Y., \& Biham, O.\ 2010, \pre, 81, 061109
\bibitem[Whittet et al.(1998)]{1998ApJ...498L.159W} Whittet, D.~C.~B., Gerakines, P.~A., Tielens, A.~G.~G.~M., et al.\ 1998, \apjl, 498, L159
\end{thebibliography}
\end{document}